\documentclass[useAMS,usegraphicx,usenatbib]{mn2e}
\usepackage{times,mathptm}   
\usepackage{balance}

\title[A fast tree for neighbour search and gravity]{A 
fast recursive coordinate bisection tree \\ 
for neighbour search and gravity}
\author[E. Gafton \& S. Rosswog]{Emanuel Gafton\thanks{E-mail: e.gafton@jacobs-university.de} \&  
Stephan Rosswog\\Jacobs University Bremen, Campus Ring 1, 28759, Bremen, Germany}

\def\msun{M$_{\odot}$}

\def\be{\begin{equation}}
\def\ee{\end{equation}}
\def\bi{\begin{itemize}}
\def\i{\item}
\def\ei{\end{itemize}}
\def\ben{\begin{enumerate}}
\def\een{\end{enumerate}}
\def\bea{\begin{eqnarray}}
\def\eea{\end{eqnarray}}

\def\edo{\end{document}}

\begin{document}

\date{Accepted 2011 July 28.  Received 2011 July 14; in original form 2011 May 12}

\pagerange{\pageref{firstpage}--\pageref{lastpage}} \pubyear{2011}

\maketitle

\label{firstpage}

\begin{abstract}
We introduce our new binary tree code for neighbour search and gravitational force 
calculations in an $N$-particle system. The tree is built in a `top-down' fashion by `recursive 
coordinate bisection'  where on each tree level we split the longest side of a cell 
through its centre of mass. This procedure continues until the average number of particles in 
the lowest tree level has dropped below a prescribed value.  To calculate the forces on 
the particles in each lowest-level cell we split the gravitational interaction into a near- and 
a far-field. Since our main intended applications are SPH simulations, we calculate the 
near-field by a direct, kernel-smoothed summation, while the far field is evaluated via a 
Cartesian Taylor expansion up to quadrupole order. Instead of applying the far-field 
approach for each particle separately, we use another Taylor expansion around the centre 
of mass of each lowest-level cell to determine the forces at the particle positions. 
Due to this `cell-cell interaction' the code performance is close to 
$\mathcal{O}(N)$ where $N$ is the number of used particles. 
We describe in detail various technicalities that ensure a low 
memory footprint and an efficient cache use.\\
In a set of benchmark tests we scrutinize our new tree and compare it to 
the `Press tree' that we have previously made ample use of.
At a slightly higher force accuracy than 
the Press tree, our tree turns out to be substantially faster and increasingly more so for 
larger particle numbers. For four million particles our tree build is faster by a factor of 25 
and the time for neighbour search and gravity is reduced by more than a factor of 6. In 
single processor tests with up to $10^{8}$ particles we confirm experimentally that the 
scaling behaviour is close to  $\mathcal{O}(N)$. The current Fortran 90 code version is 
OpenMP-parallel and scales excellently with the processor number (=24) of our test machine.
\end{abstract}

\begin{keywords}
methods: numerical -- methods: N-body simulations -- gravitation -- hydrodynamics.
\end{keywords}

\section{Introduction}
Self-gravity is a vital ingredient in the simulation of many astrophysical 
systems. Due to its long-range nature, self-gravity often becomes
the major computational burden in simulations of $N$-body systems 
or self-gravitating fluids. While many methods exist to calculate gravitational
forces under various circumstances (e.g. \citealt{hockney88}), 
for general self-gravitating systems without particular symmetries 
hierarchical methods seem to be best suited. 
The most popular such approaches are tree methods 
(\citealt{barnes86,press86,hernquist89,benz90b,couchman91,couchman95,dubinski96,stadel01,wadsley04,springel05a}; 
\citealt*{nelson09}), 
and --used to a lesser extent in astrophysics-- the fast multipole method (FMM)
originally suggested by  \cite{greengard87}. 
While tree methods
restrict the order of their inherent multipole expansion (usually to quadrupole
order) and open up further nodes for higher accuracy, the FMM instead fixes an 
acceptable error measure and then calculates whatever multipole order is necessary 
to achieve it. According to the common consensus,  tree methods are advantageous for 
problems where a moderate force accuracy can be accepted and where one 
instead invests available computational resources in larger particle numbers, while the 
FMM shows its true strength at higher accuracy. There has, however, been recent progress
in  implementations of error-controled variants of the FMM \citep{dachsel10}
and therefore the above consensus may need to come under renewed scrutiny.\\
Hybrid methods combining elements of both trees and the FMM have also been
developed and successfully tested. For example, \cite{dehnen00,dehnen02}
has developed a fast tree method that borrows from the FMM the idea of Taylor-expanding 
the fields at the `sink cells' in order to calculate the accelerations at individual 
particle positions. This `cell-cell-interaction' ensures a numerical complexity 
proportional to $\mathcal{O}(N)$ rather than
$\mathcal{O}(N \log N)$ like standard tree methods ($N$ being the number of particles).\\
Other Poisson solvers, such as the particle-particle--particle-mesh (P3M) method, 
share the same philosophy of approximating forces due to distant particles while computing
nearby interactions directly, and exhibit the same $\mathcal{O}(N \log N)$ complexity. 
Although not so widely used by the stellar astrophysics community, P3M codes have been 
successfully employed in a large variety of cosmological simulations 
\citep{efstathiou85,couchman91,macfarland98} and also combined with SPH 
\citep{evrard88}. Hybrid methods combining particle-mesh methods with trees, 
so-called TreePM methods, have also been extensively used in cosmological simulations 
\citep{xu95,bode00,bagla02,bagla03,springel05a}.\\
Here, we describe in detail our newly developed recursive coordinate 
bisection (RCB) tree. It will replace an optimized version of the Press tree 
\citep*{press86,benz89} that we have used for years in our existing codes 
\citep{rosswog02a,rosswog07c,rosswog08}. Our new tree will also become a core ingredient 
of a new SPH code that is currently under development. Since each of 
these codes has its own peculiarities in terms of input physics, time integration 
schemes etc.,  we restrict ourselves in this paper to a self-contained description
of our tree module (methods and implementation details) and defer questions such
as time integration etc. to a later point.
The main purpose of this paper is to document in detail the entity of our tree 
ingredients and their interplay for future reference. Some of our tree ingredients 
have also been used in other tree implementations, this will be discussed further in
Sec.~\ref{subsec:comparison}.\\
Our tree is designed to efficiently find neighbour particles for 
smoothed particle hydrodynamics (SPH) calculations and for the fast calculation of 
gravitational forces.  We were guided by applications which do not exhibit any particular 
symmetries, such as stellar collisions and tidal disruptions of stars by black holes.
In simulations that involve black holes particles are frequently absorbed near the horizon, 
and therefore just `repairing' an existing tree becomes difficult. Instead, such simulations
usually require a frequent  and complete re-building of the tree from scratch.
Therefore an efficient tree building phase is crucial for our intended purposes.
This is part of the reason why we decided for an RCB tree during the code design phase.\\
Our code has been written from scratch in clean Fortran 90 and does not make use of any
external libraries. It is hardware- and platform-independent and as versatile as possible,
making no assumptions about the problem type or the particle distribution. It particularly 
aims at scaling well on a large number of processors (for now using OpenMP) and at being
memory efficient. It implements a number of optimization techniques that we describe in detail
below. As we will demonstrate, our tree scales close 
to $\mathcal{O}(N)$ for large particle numbers. \\
The remainder of the paper is structured as follows. In Sec. 2 we describe in detail how 
we build and  walk the tree, and how we use it  for efficient neighbour search and gravity 
calculations. In Sec. 3 we present a set of challenging benchmark tests where we 
compare our new results against those  obtained with the  so-called `Press tree'
\citep*{press86,benz89} that is widely used in astrophysics 
(e.g. \citealt{benz90b}; \citealt*{bate95,bonnell03}; \citealt{price09,nelson09}) 
and that we have frequently used ourselves
(\citealt{rosswog99,rosswog02a,rosswog05a,price06};
\citealt*{rosswog09a}; \citealt{dan11}). In Sec. 4 we 
summarize the main features of our tree method.

\section{A Recursive Coordinate Bisection (RCB) tree}
The underlying idea of a tree method is to collect particles into hierarchically organized groups 
(sometimes we also refer to them as `nodes' or via their enclosing `cells') so 
that expensive tasks can be performed on aggregated quantities of the groups rather than
on individual particles. 
For astrophysical purposes, a tree method needs the following ingredients:
i) a strategy how to aggregate particles into a hierarchy of groups (`tree build'), ii) composition
formulae to calculate the properties of a `mother' cell from the properties of its `daughters', iii)
an `opening criterion' that decides while combing through the tree whether a node can be 
accepted `as is' or needs to be opened up into its lower level constituents, and finally 
iv) a prescription how to calculate forces from a list of aggregated nodes.\\
Particularly popular in astrophysics are the Barnes-Hut octree, henceforth `BH tree'
\citep{barnes86,hernquist87,turner95,springel05a,merlin10} and the mutual nearest neighbour
binary tree due to Press \citep{press86,benz89}.\\
The method of choice for constructing our tree is `recursive coordinate bisection'. 
Initially all particles are grouped in a root cell and  in subsequent steps the longest side of 
each cell is split through its centre of mass. This leaves an approximate balance between the 
particle numbers in each of the two resulting daughter cells. The process is repeated until 
the average particle number per cell has dropped below some pre-defined, empirically-optimised 
limit, that we will later refer to as $\overline{N}_{ll}$. Our procedure results in a binary 
tree structure that is extremely fast to build up. 
Such trees are frequently used in computer science and are often referred
to as $k$d-trees \citep{bentley75}. To our knowledge they have only been used in 
astrophysics in the {\sc pkdgrav} code \citep{stadel01} which later became the basis of
{\sc gasoline} \citep{wadsley04}.
In the way we build up our tree it delivers an adaptive mesh structure tailored to the particle 
distribution. The cells' labels carry information about local proximity so that the tree structure 
can be naturally and efficiently used to sort particles according to their spatial distribution. 
Similar techniques such as Peano or Morton space-filling curves are often used in particle 
methods to improve the computational speed via enhanced cache-coherence.\\
In the remainder of this section we describe in detail how we build up our tree 
(Sec. \ref{subsec:build}), how we traverse it (Sec. \ref{subsec:treewalk})
and search for neighbours (Sec. \ref{subsec:nei}). The gravity calculations are
described in Sec. \ref{subsec:grav} and in Sec. \ref{subsec:comparison} we compare
our approach with other work.

\subsection{Tree build} \label{subsec:build}
\subsubsection{Strategy} A tree build can either follow a `top-down' strategy where a 
root node that contains all particles is successively tesselated into
 smaller entities (`nodes' or `cells'), or a `bottom-up' approach, where, starting
 on the level of individual particles groups are formed that are subsequently
 collected into groups of groups and so on. The BH tree is an example of a
 top-down tree, the Press tree is a bottom-up tree. Building a tree in a 
top-down approach is substantially faster \citep{makino90}, but it may generate 
situations in which particles that are well separated in space are artificially placed 
in the same node, while nearby particles are placed in different nodes. The factors that 
minimize the chances of this happening are the degree of decomposition and the load 
balancing algorithm, which together define the prescription how a node is split into
its children.

\subsubsection{Degree of decomposition} 
Octrees (quadtrees in 2D) split a cell into eight (four in 2D) daughter cells, binary trees
only in two. Therefore octrees are $\frac{\log 8}{\log 2}=3$ times shorter than 
binary trees and usually faster to construct. But by forcing all sides to be 
split, they tend to create elongated daughter cells out of elongated parent cells.
This is usually not desirable since such cells can have large high-order multipole
moments and can thus introduce large truncation errors in the force calculation.
In our binary tree approach we divide a node through its longest edge, which tends
to drive daughter cells closer to the desired, compact shape. Binary trees also 
tend to return shorter interaction lists for a given accuracy 
\citep{anderson99,waltz02}, which reduces the number of direct force evaluations 
and thus speeds up the gravity calculations.

\begin{figure}
\centering
\includegraphics[clip,width=84mm]{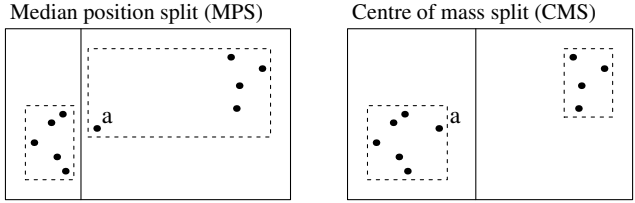}
\caption{A 2D illustration of two ways to split a given cell: the median position split (MPS, left)
and the centre of mass split (CMS, right). The MPS leads to perfectly balanced 
daughter nodes, but can artificially separate particles that naturally belong to the same group,
as illustrated by particle $a$. Since we calculate our gravitational
forces by a Taylor expansion around the centre of mass, the MPS can lead to large force errors,
therefore we prefer the CMS for building our tree.}
\label{fig:loadbalancing}
\end{figure}

\subsubsection{Load balancing} 
\label{subsec:splitting}
An important concern is keeping the decomposition of a tree
balanced in the sense that sibling nodes contain comparable amounts of particles.
Most implementations of the BH tree are {\it spatially-balanced}, cutting the edge 
of a cell through its middle. While this method is extremely fast because one only 
has to compute the average of two coordinates, it can lead to very uneven particle 
distributions. {\it Density-balanced} trees, on the other hand, ensure that comparable 
computational domains remain on each side after the cut. In the design phase of our 
RCB tree, we have experimented with a median position split (MPS) and a centre of 
mass split (CMS), see Fig.~\ref{fig:loadbalancing}. The MPS produces perfectly balanced
particle numbers in each daughter cell, but it can lead to an artificial separation 
of particles that one would consider as belonging to the same group into different nodes.
This is more than just an aesthetic flaw, since due to our Taylor expansion around the
cell centre of mass, see Sec. \ref{subsec:grav}, particles that are artificially
separated from their natural group of peers (e.g. particle $a$ in
Fig.~\ref{fig:loadbalancing}, left) suffer substantially larger truncation errors for the 
MPS than for the CMS case. In our experiments we found that the relative force accuracy 
of such `renegades' can be improved by orders of magnitude if the CMS is used instead.
Another example would be a stellar binary system containing different numbers of 
particles per star. The CMS assigns each star its own node, while the MPS would 
place some of the particles from the heavier star in the node of the lighter star, 
creating elongated cells and substantially larger multipole truncation errors.\\
In passing we note that one does not necessarily have to split a cell through
its centre of mass, although this is convenient for our applications. 
If systems with charges of different signs are simulated, say in plasma physics,
it may be more convenient to split through the `centre of number' (by simply
replacing the particle masses by a weight of `1'), and then use the 
proper charges for the calculation of the needed multipoles. \\
By choosing the CMS rather than the MPS we pay the price that empty lowest-level cells
(`ll-cells') can, at least in principle, occur. The CMS procedure will often assign 
different numbers of particles to sibling cells and therefore one can eventually end 
up with empty ll-cells at the expense of other ll-cells being over-populated. The 
latter will be very compact (since high densities are the reason why imbalanced 
distributions are obtained in the first place), and the multipole expansion will 
converge. At the same time, empty cells (less than 1\%
of the total number of cells even for pathological cases) can be safely ignored, 
since they do not contain neighbours and do not contribute to gravity either.

\subsubsection{Particle sorting} 
One technique that is frequently used in particle methods in order to reduce
the number of cache misses is re-ordering the particles in memory based on
their physical proximity (e.g. \citealt{springel05a,thacker06,nelson09}). In experiments
of stellar collisions we had found that particle sorting can easily improve
overall runtimes by a factor of a few, concluding that the overhead
due to sorting is certainly worthwhile the extra effort.
The RCB tree uses a customised partitioning 
algorithm inspired by the quick-sort method \citep{hoare62} to sort 
the particle arrays.\\
We first compute the centre of mass of
a node, then decide which direction $x_{i}$ to split, and subsequently perform
a single iteration through all the particles in the node. During this 
iteration, we move all particles that have the $x_{i}$ coordinate 
smaller than that of the centre of mass at the beginning 
of the array, and all particles with larger $x_{i}$ at the end of the 
array. The boundary between these two subsets then tells us where the 
particles of the left child node end and those of the right child node 
begin. There is no need to sort these two subsets of particles any further because 
on the next level one might have to split along a different side.\\
One advantage of sorting the particles is that it eliminates the need 
to keep a list of all the particles in an ll-cell, the indices of the 
first and the last particles being sufficient (this actually holds true 
for any node, since particles are `sorted' on each level). In addition, 
looping through the particles in any one node requires no jumps through 
the memory, and hence the array operations (such as the computation of 
the centre of mass) prove to be extremely cache-efficient.\\
It is known that typical quicksort implementations have a worst-case complexity
of $\mathcal{O}(N^2)$ \citep[Sec.~8.2]{press92}, usually occuring if one  
repeatedly uses poorly-chosen pivots on already-sorted arrays. In our case,
we compute the centre of mass anyway (since we later need it in the gravity
calculations) so the pivot element is always optimally chosen. Since 
on each of the $\log N$ levels we traverse the arrays just once ($\mathcal{O}(N)$), the
complexity of this subroutine is always $\mathcal{O}(N\log N)$.
As will be shown below, see 
Fig.~\ref{fig:TimingsRCBLogarithmic}, the tree build takes just a small
fraction of the CPU-time (less than a percent) in comparison 
to neighbour search and gravity, and the sorting, in turn, is just a small
fraction of the tree build. The investment of time in sorting
is completely negligible but speeds up the tree build subroutine
substantially. We also gain nearly a factor three
by substantially enhanced cache access during force evaluations.

\begin{figure}
\centering
\includegraphics[clip,width=84mm]{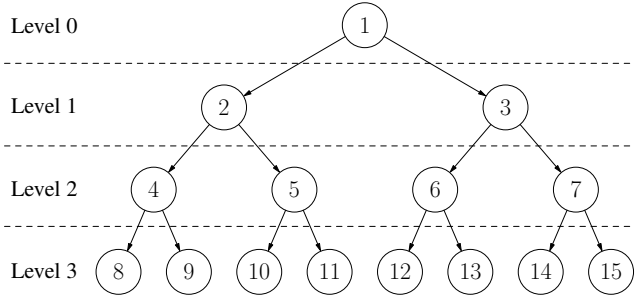}
\caption{Node labelling convention in the RCB tree.
Many useful relations can be trivially recovered by simple arithmetic
operations on the node labels without any need for additional memory consumption.
See main text for details.}
\label{fig:nodelabelling}
\end{figure}

\subsubsection{Node labelling} \label{sec:labelling}
A well-chosen node labelling system eliminates the 
need for pointers that link parent nodes and their children. In the RCB 
tree, instead of storing the nodes as the components of a linked list 
\citep{dubinski96,dehnen02,nelson09}, we simply use a one-dimensional 
array, with the relationship between the nodes encoded in their indices.\\
Our node labelling conventions are illustrated in Fig.~\ref{fig:nodelabelling}.
For a tree with $k$ levels the total number of nodes is $2^k-1$. The 
best choice is to assign the root node the label `1', place it on level 
`0', and then continue to increment the labels level-by-level. With this
labelling convention, simple `index gymnastics' allows to recover various
node relations without additional memory consumption. For example:
\ben
\renewcommand{\theenumi}{(\arabic{enumi})}
\i the children of node $n$ are $2n$ and $2n+1$;
\i the level of cell $n$ is $[\log_2 n]$ (Gau{\ss} bracket);
\i the first node on level $k$ is $2^k$;
\i the last node on level $k$ is $2^{k+1}-1$;
\i the number of cells on level $k$ is $2^k$;
\i the total number of descendants of a cell on 
   level $k$ is $2^p-2$, where $p$ is the number of levels greater than 
   or equal to $k$. 
\een
\begin{figure}
\centering
\includegraphics[clip,width=75mm]{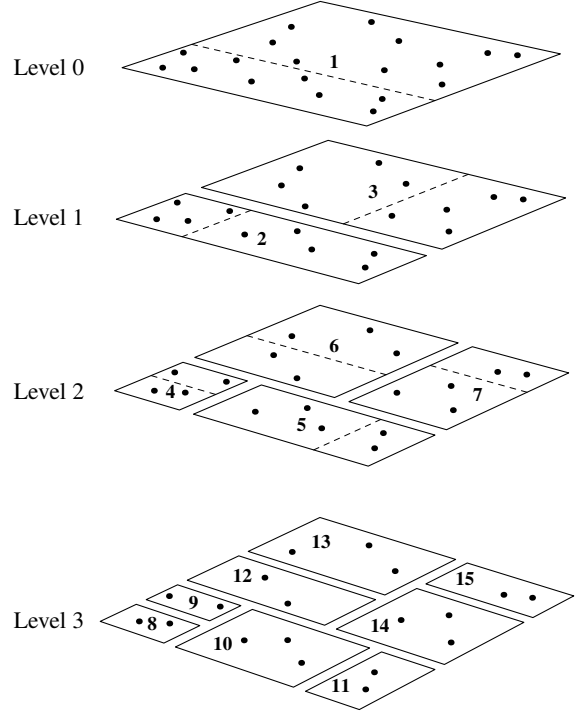}
\caption{Illustration of the spatial splitting and labelling
of cells on different levels for a two-dimensional particle 
distribution. Dashed lines (planes in 3D) are perpendicular
to the longest cell side and cross the centres of mass. 
The tesselation ends once the average number of particles per 
lowest-level cell has dropped below a prescribed value, $\overline{N}_{ll}$.}
\label{fig:treelevels3D}
\end{figure}
The last relation is used, for example, when entire branches are
discarded during the tree walk. Fig.~\ref{fig:treelevels3D} illustrates
how the node labels relate to the spatial tesselation for a
two-dimensional particle distribution.

\subsubsection{Multipole moments} 
During the tree building phase, a number of physical 
quantities are calculated and stored for each node: position of the lowest-left 
corner, size, coordinates of the geometrical centre and of the centre of mass, 
and the multipole moments. For the gravitational forces from distant nodes
we use a simple Cartesian multipole expansion up
to quadrupole order. Since the dipole moment vanishes when the origin of 
the multipole expansion coincides with the centre of mass, one only needs
to store the monopole and the quadrupole moments. The monopole is simply the 
mass of all the particles in the cell and can be computed very fast with the 
\texttt{SUM} command, since all particles of a given node, having been sorted, 
are contiguous in memory. For the ll-cells one computes the components
of the traceless quadrupole tensor due to each individual 
particle $a$,
\be\label{eq:multipole_cartesian}
Q_{ij}^a=3 m^{a} \Delta x_{i}^{a} \Delta x_{j}^{a}-m^{a}( \Delta r^{a})^{2}\delta_{ij}^{a},
\ee
where $m_a$ is the particle mass, $\Delta x_k^a$ are the
components of the particle $a$'s offset from the cell centre of mass and 
($\Delta r^{a})^2= \sum_i (\Delta x_i^a)^2$.
Subsequently, these quadrupole moments are `shifted up' to the parent nodes. 
Specifically, the quadrupole moments $Q_{ij}^{P}$ of a parent node $P$ are 
computed in terms of the quadrupole moments $Q_{ij}^{C}$ of its children $C$ as:
\be\label{eq:multipole_shifted}
Q_{ij}^{P}= \sum_{C}\left[Q_{ij}^{C}+3M^{C}\Delta x_{i}^{C} \Delta x_{j}^{C}
            - M^{C}(\Delta r^{C})^{2} \delta_{ij}\right],
\ee 
where $M^C$ is the mass of child $C$, $\Delta x_{i}^{C}$ is 
the distance between the centres of mass of the parent and of its child
along the axis $x_{i}$, and $\Delta r^{C}$ is the 
distance between their centres of mass: $(\Delta r^{C})^2=\sum_{i} (\Delta x_{i}^C)^{2}$.

\subsubsection{Parallelisation} Since each level of the tree depends 
on the previous levels (the particles must be sorted in the 
parent node before the children nodes can be created) simple 
OpenMP work sharing constructs will not work out of the box. 
Instead, we use an MPI-like approach: assuming that 
$2^{k}$ processors are to be used for the tree build, one 
builds the tree down to level $k$ on just one processor, and 
then `distributes' each of the $2^{k}$ nodes on level $k$ 
to one processor. Since the particles are sorted down to 
level $k$, each processor can now build its own `sub-tree' 
by simply ignoring the rest of the tree and acting as if its 
assigned node were the root node.

\subsubsection{Updating the tree} Although heavily optimised and never 
taking more than 1\% of the total computational time, the 
tree build is still an expensive operation that should be 
avoided when possible. For this reason, the integrity of the 
tree is checked after every time step: if the particles have 
not moved out of their ll-cells then the tree does not need 
to be rebuilt. It is sufficient to simply update the relevant 
quantities of the ll-cells (centre of mass, radii of influence 
and quadrupole moments), and then compute these quantities for 
all the larger cells in a bottom-up manner. Since no tesselation and 
sorting are required, the procedure is much faster than a 
full-scale tree build and can be used until one particle 
crosses the border of its ll-cell. At that point, the tree could 
in principle be `repaired' by locally adjusting 
the tree structure where necessary. For now, however, we 
completely rebuild the tree whenever this `integrity check' 
fails, otherwise we just update it.

\subsection{Tree walk} \label{subsec:treewalk}
In order to find properties (such as neighbours or gravitational 
forces) of particles in the ll-cells we have to `walk the tree'. 
Recursive tree walks can be programmed elegantly 
and in a very compact way, but they are usually too inefficient 
for use in astrophysical applications, since they complicate to 
take optimization decisions at compile time and each recursive 
call comes with its own computational overhead.
\begin{figure}
\centering
\includegraphics[clip,width=84mm]{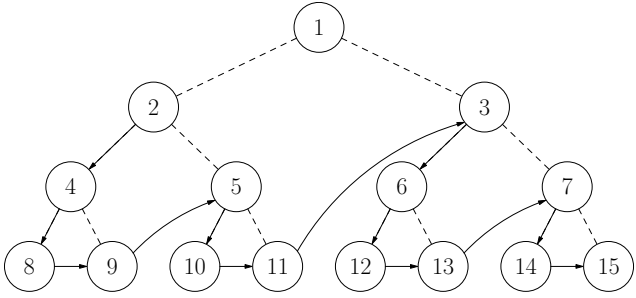}
\caption{The order in which tree nodes
are accessed during a tree walk. Jumps between 
non-neighbouring cells are indicated by curved arrows. Note 
that the original connection between a parent cell and its 
right daughter, marked with dashed lines, is no longer 
significant for the tree walk. All tree walks begin with 
node 2.}
\label{fig:dfswalk}
\end{figure}
We have implemented a modified version of the depth-first search algorithm 
\cite[\S 22.3]{cormen01} which has been used in astrophysical 
applications, for example, by \cite{dubinski96}. The order in which 
the tree nodes are accessed is shown in Fig. \ref{fig:dfswalk}.\\
A tree walk requires a criterion (one for a neighbour and one
for a gravity walk) for the decision whether to accept a node 
or to open it and examine the two daughter nodes. If an ll-cell 
cannot be accepted then all its particles are processed individually. 
A tree walk ends when all the nodes in the tree have been either 
opened or skipped. Since the root node contains all the particles, 
it can never be accepted by any criterion; hence, the tree walk 
always begins with node 2.

\subsubsection{Tree vector} 
\begin{figure}
\centering
\includegraphics[clip,width=84mm]{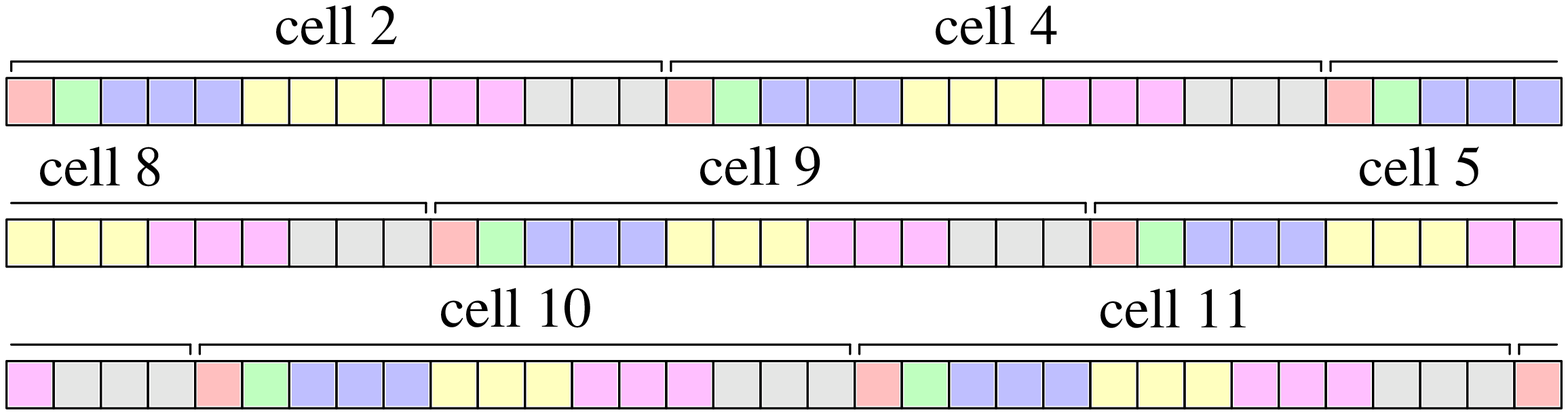}
\caption{The `tree vector' is a one-dimensional 
array which stores all the relevant properties of all nodes 
contiguously, in exactly the order in which they are accessed during 
the tree walk, see Fig. \ref{fig:dfswalk}. This allows for a very
cache-efficient data access. In a typical simulation with 
a few million particles, the use of the tree vector can reduce 
the time spent for tree walks by as much as two orders of magnitude.}
\label{fig:treevector}
\end{figure}

The tree needs to be walked at least once
per time step for every ll-cell that needs an update, therefore
the tree walk efficiency is of key importance for the overall runtime.
The computations involved in the neighbour search (simple comparisons) 
are not complicated, but the sheer memory span that the algorithm 
has to traverse (a tree can have millions of cells scattered throughout 
the RAM) can slow down the subroutine considerably. Therefore just accessing
the node data in the order of the tree walk would lead to continuous and repeated
cache misses for every tree walk and therefore would cause a serious slowdown. 
The tree data, however, is always accessed in the same order.
This led us to the idea of introducing a `tree vector', a one-dimensional 
array which stores all the needed data (label, level, centre of mass, geometrical 
centre, position, size) of the nodes contiguously, in exactly the order 
in which they are accessed during the tree walk. This is illustrated in
Fig.~\ref{fig:treevector}. 
The tree vector is created once per time step, 
directly after the tree build, and it is subsequently used in all the tree walks. 
Inefficient data access therefore just occurs once (in the tree vector creating 
stage) rather than hundreds of thousand times during repeated tree walks.
The subsequent tree walks all receive their data  as a contiguous batch 
of elements from the tree vector. If a node has to be opened, the next batch 
of elements, corresponding to 
the node's left daughter, is read from the tree vector, and the code simply 
advances a counter; usually, chances are that this batch is already loaded in 
cache (since whole segments, and not individual numbers, are read from the 
RAM at once). If the cell has to be skipped then all its $2^{p}-2$ descendants,
see Sec.~\ref{sec:labelling}, are skipped, in which case the tree vector 
counter is simply incremented by the required amount. The tree walk ends when 
the counter reaches the end of the tree vector. In our benchmark tests
with a few million particles distributed according to a Sobol sequence,
see Sec. \ref{sec:benchmark}, the use of a tree vector speeds up the
tree walks by as much as two orders of magnitude.

\subsection{Neighbour search}
 \label{subsec:nei}
In SPH, continuous or `smoothed' quantities at a given position are 
obtained by summing up the kernel-weighted properties of contributing 
particles. Similarly, gradients 
of fluid properties are calculated as sums over the analytically known 
kernel gradients (see, e.g., \cite{rosswog09b} for a recent review of the SPH method).
We use here the `standard' cubic spline kernel of \cite{monaghan85a}, 
which has compact support and vanishes outside of a finite radius $2h$. 
Since contributing particles (`neighbours') need to be known at 
every time step, it is crucial to have a very efficient algorithm to identify 
them.

\subsubsection{Neighbour walk}
\begin{figure}
\centering
\includegraphics[clip,width=78mm]{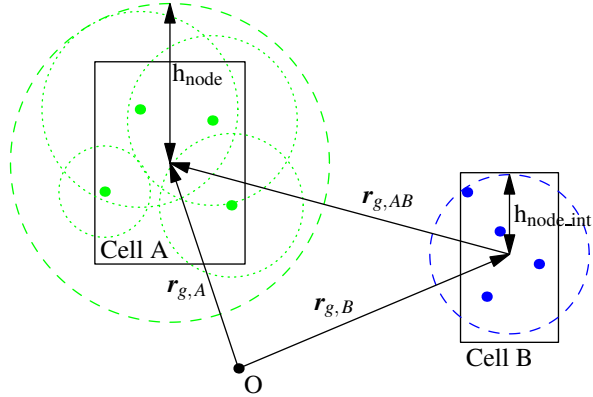}
\caption{The radius $h_{\mathrm{node}}$ of cell $A$ (dashed green circle) 
encompasses the kernels of all the particles in $A$ (dotted green circles). 
The interaction radius $h_{\mathrm{node\_int}}$ of cell $B$ (dashed blue circle) 
encompasses all particles in $B$. If we denote the distance vector between the 
geometric centres of the cells by $\mathbfit{r}_{g,AB}=\mathbfit{r}_{g,A}-\mathbfit{r}_{g,B}$, the 
`reachability condition' translates into $|\mathbfit{r}_{g,AB}| < h_{\mathrm{node}}(A)+h_{\mathrm{node\_int}}(B)$. 
If this test fails, cell $B$ is discarded and all its children are skipped in the tree walk.}
\label{fig:neighbours}
\end{figure}
To determine possible neighbour candidates, we walk the tree as described above
for every ll-cell that requires an update. 
During the tree walk we check whether a candidate cell $B$ can be reached from
the cell of interest, cell $A$. To this end we had assigned in the tree building phase
to each ll-cell the `interaction radius' $h_{\rm node}(A)$, which encloses the 
volume covered by the individual kernels of all its particles, see Fig.~\ref{fig:neighbours}. 
Moreover, each cell $B$ had been assigned a radius $h_{\rm node\_int}(B)$ that 
starting from the geometric centre of cell $B$ enclosed all its particles.  The particle 
content of cell $B$ is included in a `candidate list' for possible neighbours of 
particles in cell $A$, if
\be
|\mathbfit{r}_{g,AB}|\equiv|\mathbfit{r}_{g,A}-\mathbfit{r}_{g,B}| < h_{\mathrm{node}}(A)+h_{\mathrm{node\_int}}(B),
\ee
i.e. if particles in cell $A$ can possibly reach at least one of the particles in cell $B$.
Although not strictly necessary, in a subsequent step we cull the candidate list 
to the true neighbours to save some computational effort during the later summation 
stage (to calculate densities and the hydrodynamic derivatives).

\subsubsection{Neighbour list} 
Storing a list of neighbours for each particle can be 
problematic: we typically use a neighbour number of $\sim 100$ 
per particle \citep[Sec. 2.9]{rosswog09b}, but under certain circumstances 
individual particles can have a much larger number of neighbours. This can happen,
for example, if low-density regions with large smoothing lengths interact with 
high-density regions with small smoothing lengths. If a low density particle is 
expanding, even a small fractional increase of its smoothing length can lead to 
a large jump in the neighbour number if suddenly the high density region `becomes 
visible'. Under such circumstances one has two possibilities:
either fine-tune the smoothing length immediately to a value so that the neighbour 
number is in an acceptable range, which can introduce a substantial amount of 
numerical noise, or accept the large neighbour number for a few time steps 
for this particle, in which case one has to take care of the storage of a very long 
neighbour list for the respective particle. Since such sitations are exceptions, it is a waste
of memory to store `worst-case-sized' lists of neighbours in a fixed bi-dimensional array 
(such that neighbours $1...k$ of particle $n$ are stored at positions \texttt{(n,1:k)}); 
one would have to allocate the maximum space for each particle, even though most of 
them will have far fewer neighbours.\\
An elegant solution to this problem is to store the neighbours in a one-dimensional list. 
This comes with a slight computational overhead, since one has to also store, 
for each particle, the total number of neighbours and the index of its first 
neighbour in the one-dimensional list. The memory savings, on the other hand, 
are tremendous, for one has to only allocate an `average' number of neighbours 
per particle, rather than a `maximum' one.

\subsubsection{Parallelisation} 
If the ll-cells whose neighbours have to be updated are 
distributed to different processors, the computations are independent and can be 
programmed with OpenMP work-sharing constructs. We use dynamical scheduling -- 
gradual allocation of the iterations to various threads as they become idle --
in order to prevent load imbalance. Combining the results in a one-dimensional 
array, however, constitutes a possible bottleneck, since different threads 
might want to simultaneously modify the same array, in which case they are 
queued in the order in which they finish their calculations. This introduces 
a waiting time of a few percent of the total neighbour search, but is 
nevertheless preferable to using a bi-dimensional array.

\subsection{Gravitational forces} \label{subsec:grav} \nocite{price07d}
To achieve accurate gravitational forces at a moderate computational effort, 
we split gravity into a near- and a far-field contribution. The near-field 
component is crucial for the accuracy and it is obtained by a kernel-smoothed
direct summation, while the far-field contribution is calculated via a low-order
multipole expansion up to quadrupole order. The gravitational 
constant $G$ is set to unity throughout the paper.

\subsubsection{Multipole acceptance criterion} \label{subsubsec:MAC}
During a gravity tree walk a multipole acceptance criterion (MAC) decides
whether a node can be accepted with its multipole moments or whether it needs to be 
further resolved into its constituents for higher accuracy. In the first case 
all descendants can be skipped, in the second case the tree walk advances down 
the tree until either the acceptance criterion is met, or a lowest-level cell is reached, 
in which case all its particles are added to a near-field list. After experiments 
with various criteria that have been published in the literature 
\citep{salmon94,nelson09} we decided for the simple geometric MAC introduced 
by \cite{barnes86}: a cell $B$ is accepted if the opening angle under which it 
is seen drops below a prescribed accuracy parameter $\theta$: 
\be\label{geoMAC}
\frac{H_B}{R_{AB}} < \theta \quad {\rm with} \quad H_B= \mathrm{MAX}(S_{i,B}).
\label{eq:MAC}
\ee Here, the $S_{i,B}$ are the side lengths of cell $B$ and 
$R_{AB}= |\mathbfit{r}_{A}-\mathbfit{r}_{B}|$ is the distance between the centres of mass of
cells $A$ and $B$. For $\theta=0$ this reproduces, of course, the direct 
summation case. 
Furthermore, if a cell $B$ contains potential neighbour particles,
we add its constituents to the near-field list,
and do not accept the cell, independent of the MAC.
This leads to relatively long near-field lists ($\sim 800$ particles), which 
come at an $N^2$-price, but which are also important for high force accuracy. Moreover, 
the near-field contribution parallelizes perfectly well, so for large parallel calculations
this becomes an acceptable procedure. As we will show below, our algorithm is
very efficient despite the relatively long near-field lists. The rare, 
potentially unbounded errors introduced by the simple geometric MAC and
described by \cite{salmon94} are avoided in our code, 
as explained in Sec. \ref{subsec:accuracy}.

\subsubsection{Far field} 
The multipole expansion of the gravitational potential $\Phi(\mathbfit{r})$
due to a distant node with mass $M$ and quadrupole moments $Q_{ij}$ reads in 
Cartesian coordinates:
\be\label{eq:multipole_potential}
\Phi(\mathbfit{r})=-\frac{M}{r}-\frac{1}{2}\sum_{i,j}\frac{x_{i}x_{j}}{r^{5}}Q_{ij} + \mathcal{O}(r^{-7}).
\ee 
\begin{figure}
\centering
\includegraphics[clip,width=84mm]{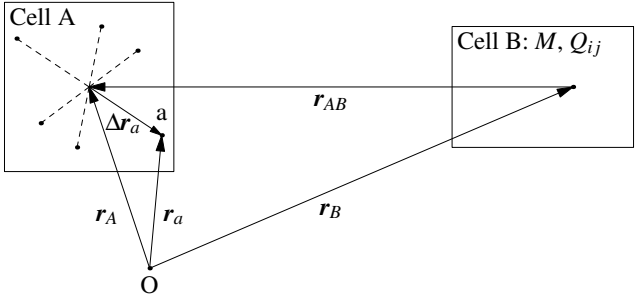}
\caption{Forces at particle positions are calculated via a Taylor
expansion around the cell centres of mass. For a particle $a$ at position $\mathbfit{r}_{a}$,
the acceleration $\mathbfit{f}_{a}$ due to cell $B$ is obtained by Taylor expanding
$\mathbfit{f}(\mathbfit{r}_{A}+\Delta\mathbfit{r}_{a})$ around $\mathbfit{r}_{A}$.}
\label{fig:localtaylor}
\end{figure}
Once a list of acceptable, distant nodes has been identified via a tree walk, the resulting 
accelerations can be calculated as the gradients of the truncated potential, 
$\mathbfit{f}(\mathbfit{r})=-\nabla\Phi(\mathbfit{r})$.  We calculate the far gravity acceleration 
on particles in a given ll-cell via `cell-cell interaction', similar to what is done in the FMM 
and the tree code of \cite{dehnen00,dehnen02}. To this end we calculate the Taylor
expansion of the forces around the centre of mass of an ll-cell, see Fig. \ref{fig:localtaylor}. 
For a particle $a$ in cell $A$ the far-gravity contribution reads to second-order:
\be\label{eq:ll_cell_taylor}
\mathbfit{f}_{\rm fg}(\mathbfit{r}_a)= \mathbfit{f}_{\rm fg}(\mathbfit{r}_A+\Delta\,\mathbfit{r}_a)
            \simeq \mathbfit{f}_{\rm fg}(\mathbfit{r}_A)+\mathbfss{J}_A \Delta\,\mathbfit{r}_a+
\frac{1}{2}\Delta\,\mathbfit{r}_a^{\mathsf{T}}\mathbfss{H}_A  \Delta\,\mathbfit{r}_a.
\ee 
Here, $\mathbfss{J}_A$ and $\mathbfss{H}_A$ are the Jacobian and Hessian as 
evaluated at point $\mathbfit{r}_A$.  In three dimensions they only have 6 and 10 
independent components, respectively, due to the equality of mixed partial derivatives. 
This Taylor expansion allows to evaluate the far gravity only once per ll-cell and 
ensures the approximate $\mathcal{O}(N)$ behaviour for large $N$.

\subsubsection{Near field} 
Our goal is to simulate a self-gravitating fluid rather than a point
particle system. Therefore, the mutual interaction between nearby particles 
needs to be softened. We use  the SPH
kernel function $W$ to smooth both hydrodynamics and near-field gravity in
a consistent way. The kernel-softened force 
\citep[e.g.][]{dyer93,dehnen01}
on a particle $a$ due to $b$ is given by:
\be
\mathbfit{F}_a= - \frac{m_a m_b}{r_{ab}^2} \eta_{ab} \hat{e}_{ab},
\ee 
where the softening is mediated via
\be
\eta_{ab}= 4 \pi \int_0^{r_{ab}}  r^2 W(r,h) dr.
\ee
Here we have used $\hat{e}_{ab}=\mathbfit{r}_{ab}/r_{ab}$,
$\mathbfit{r}_{ab}= \mathbfit{r}_{a} - \mathbfit{r}_{b}$. 
Thus the gravitational force is smoothly switched off as the particles approach
each other. For constant $h$, Poisson's equation shows that this force law 
corresponds to the force on a point particle $a$ due to a density 
$\rho(\mathbfit{r}_a)= m_b W(r_{ab},h)$. For the commonly used cubic spline kernel
\citep{monaghan85a} the near gravity acceleration on particle $a$ then becomes
\be
\mathbfit{f}_{\rm ng}(\mathbfit{r}_{a}) = -\sum_{b\neq a} m_{b} 
S(r_{ab},h) \hat{e}_{ab}
\label{eq:direct_kernel_sum}
\ee 
with
\be
\label{kernel_gravity}
S(q)=\left\{
\begin{array}{ll}
1/h^2\left(\frac{4}{3}q - \frac{6}{5}q^3 + \frac{1}{2}q^4\right), & 0 \leq q < 1 \\
1/h^2\left(\frac{8}{3}q - 3q^2 + \frac{6}{5}q^3 - \frac{1}{6}q^4 - \frac{1}{15q^2}\right), & 1 \leq q < 2 \\
1/r^2, & q \geq 2
\end{array}
\right.
\ee
and $q= r_{ab}/h$.
To ensure exact conservation, $S$ should be symmetric with respect to the particle 
labels $a$ and $b$, which we enforce by using  $h= 0.5(h_a+h_b)$.\\
Alternatively, a discretized set of self-gravitating SPH equations
can be derived consistently via a variational principle from a Lagrangian which
contains an ideal fluid contribution and additional self-gravity terms \citep{price07a}. 
Such an approach delivers consistently softened SPH equations and in addition also
`gravitational grad-$h$' terms which are similar to the correction terms derived for the 
SPH equations \citep{springel02,monaghan02}. Such a treatment is beyond our current focus,
but if desired, it can be implemented in a straight forward way.

\subsubsection{Parallelisation} 
Since computing the gravitational accelerations acting on two different ll-cells involves unrelated 
calculations, dynamically scheduled OpenMP work-sharing constructs work well for the gravity calculations and no 
specific optimisation is needed.

\subsection{Comparison with other work} \label{subsec:comparison}
Some elements of our RCB tree have been used in previous work.
Here we will briefly summarize similarities and differences.
The most commonly used type of tree in astrophysics is the Barnes-Hut oct-tree. 
Since we have outlined some basics above and since this type of tree
is fundamentally different from ours, we do not further discuss it here.\\
The only astrophysical tree that is built similar to ours (as a $k$d-tree), is the one used in
{\sc gasoline} \citep{wadsley04} which is based on the {\sc pkdgrav} code \citep{stadel01}. 
It also does not build the tree down to the particle level,
but instead down to what the authors call `buckets', which correspond to our ll-cells. There
are a number of differences between {\sc gasoline} and our tree. In {\sc gasoline}
cells are split according to MPS while we use CMS, see Sec.~\ref{subsec:splitting}. We 
found the CMS splitting to be substantially more accurate in terms of
worst case errors. 
The authors calculate far-gravity contributions {\em per particle} using a hexadecupole expansion
for acceptable cells while we calculate it {\em per ll-cell} using a quadrupole expansion
and obtain the forces at particle positions via a Taylor expansion. Also the  near-gravity
approach differs. They base their direct summation list decision on a purely cell 
property-based opening criterion, their Eq. (1), which does {\em not} involve any 
information about the smoothing lengths. Therefore,
no consistency between hydrodynamics and gravity is guaranteed in the sense that hydrodynamically
interacting particles also interact (i.e. are smoothed) gravitationally. In principle, this can deteriorate
the conservation properties, but depending on the chosen parameters this may or may not be relevant in
practical simulations.
We pursue a more conservative (though somewhat more costly) strategy in the sense that reachable cells
are added to the direct summation list, independently of the MAC. However, if further optimization is desired,
this could be changed by simply modifying one line in the gravity-walk subroutine.\\
As will be shown below, our tree scales close $\mathcal{O}(N)$, which is due to the far-force being
calculated via a Taylor expansion only once per ll-cell. Such cell-cell interactions are at the heart of
the FMM method \citep{greengard87}. 
The idea of cell-cell interactions, i.e. using Taylor expansions at both the `sink' and the `source'
side, has been used in the $\mathcal{O}(N)$ tree of Dehnen \citep{dehnen00,dehnen02}. 
Contrary to our approach, Dehnen
uses a standard oct-tree structure and makes explicit use of symmetries in the double-sided Taylor expansions
in order to reduce the operational effort. Combined with a MAC that is symmetric in the properties of the two
interacting cells this explicitely enforces momentum conservation by ensuring that Newton's third law is obeyed.
The benefits of this approach come at the price of larger memory consumption; also, the explicit use of the 
symmetries in Taylor expansion coefficents (cells are evaluated as both `sinks' and `sources')
prevents the use of standard tree walks. Instead, Dehnen uses a two step approach with an `interaction' and 
`evaluation' phase, see Sec. 3.2 in \cite{dehnen02} for details. The benefit of exact conservation, however,
needs to be given up when individual time steps are used.\\
Moreover, Dehnen's tree is designed for N-body codes rather than for simulating self-gravitating fluids
as in our case. Therefore, the MAC is not `overruled' by a `reachability' criterion as in our (conservative)
approach described in Sec.~\ref{subsubsec:MAC}. As an effect, the near-field lists can be kept shorter which
reduces the cost of the direct summation part of the code. As outlined above, such an approach could be easily
implemented in our tree, but for now we stick to the more conservative (and slightly more expensive) approach.

\section{Benchmark results} \label{sec:benchmark}
\begin{figure*}
\begin{minipage}[t]{0.31\linewidth}
\centering
\includegraphics[clip,angle=270,totalheight=50mm]{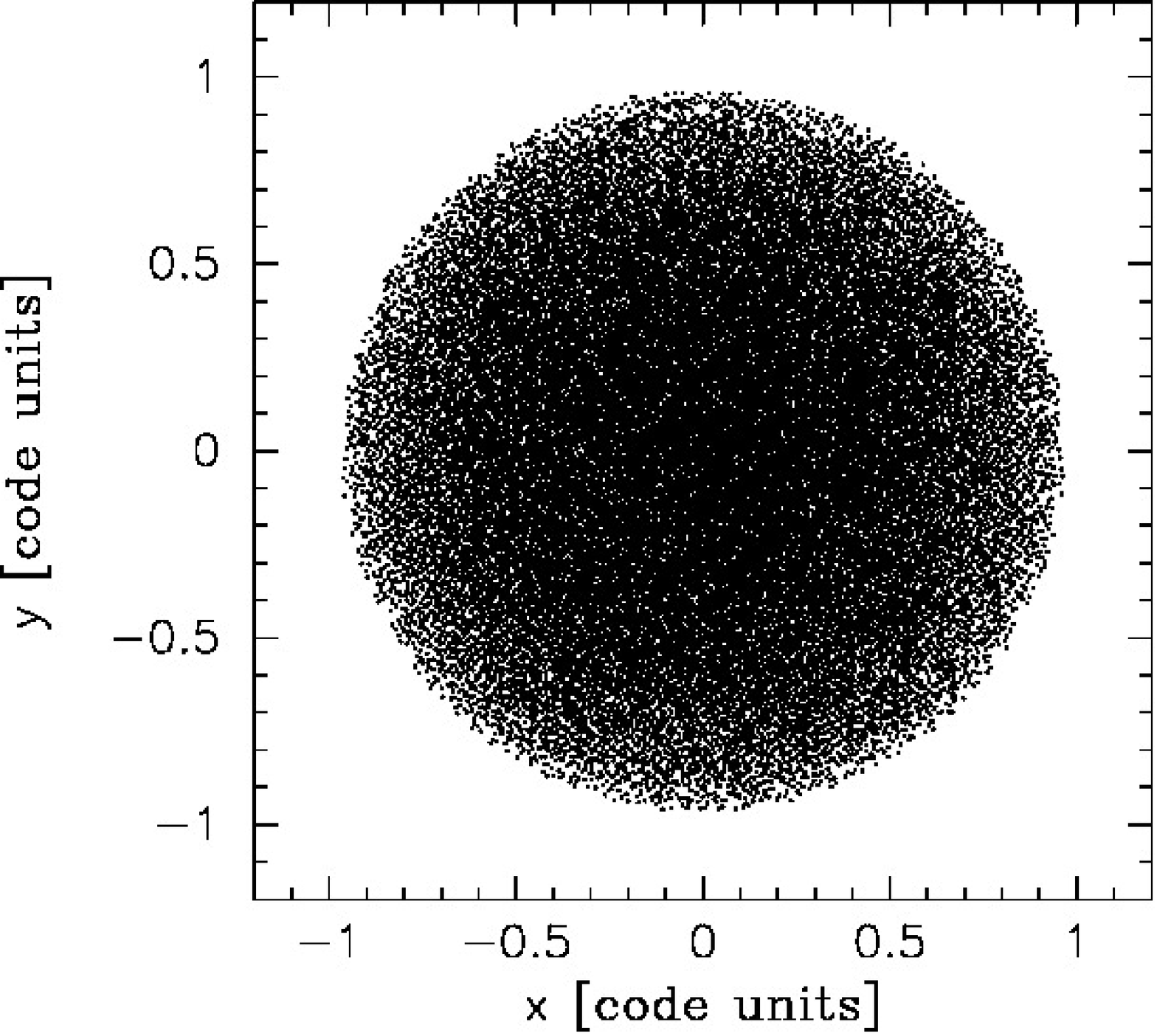}
\caption{A particle distribution within a sphere, obtained via a Sobol quasi-random sequence. Such a particle
distribution is homogeneous, but --due to its randomness-- represents a `worst case scenario' for efficient
memory access.}
\label{fig:sobol_splash}
\end{minipage}
\hspace{3mm}
\begin{minipage}[t]{0.31\linewidth}
\centering
\includegraphics[clip,angle=270,totalheight=50mm]{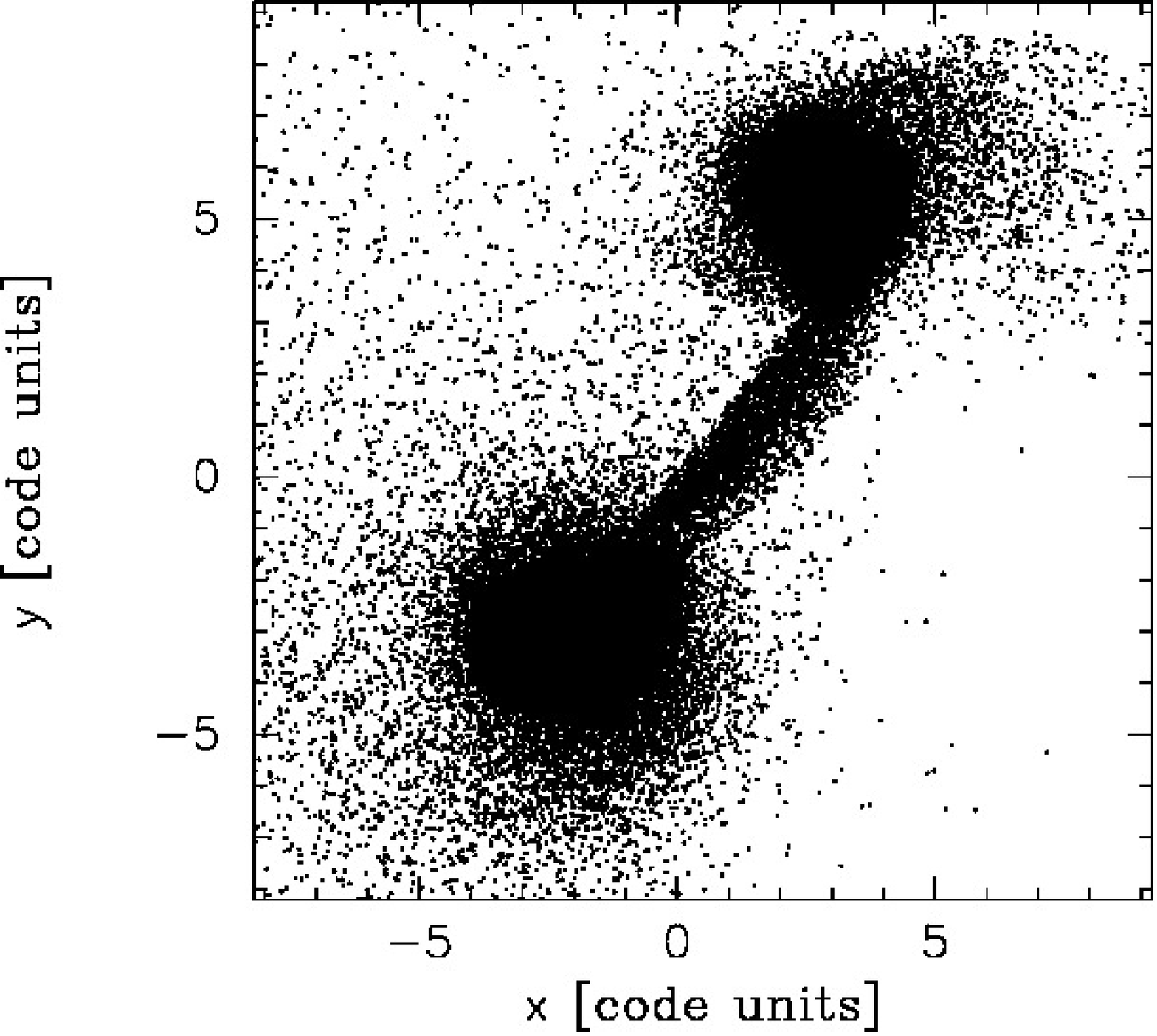}
\caption{Snapshot from a white dwarf -- white dwarf off-centre collision (602506 particles). 
The smoothing lengths span a range of over three orders of magnitude, due to the large density 
discrepancy between the stellar centers and the accretion flow between them.}
\label{fig:wdwd_splash}
\end{minipage}
\hspace{3mm}
\begin{minipage}[t]{0.31\linewidth}
\centering
\includegraphics[clip,angle=270,totalheight=50mm]{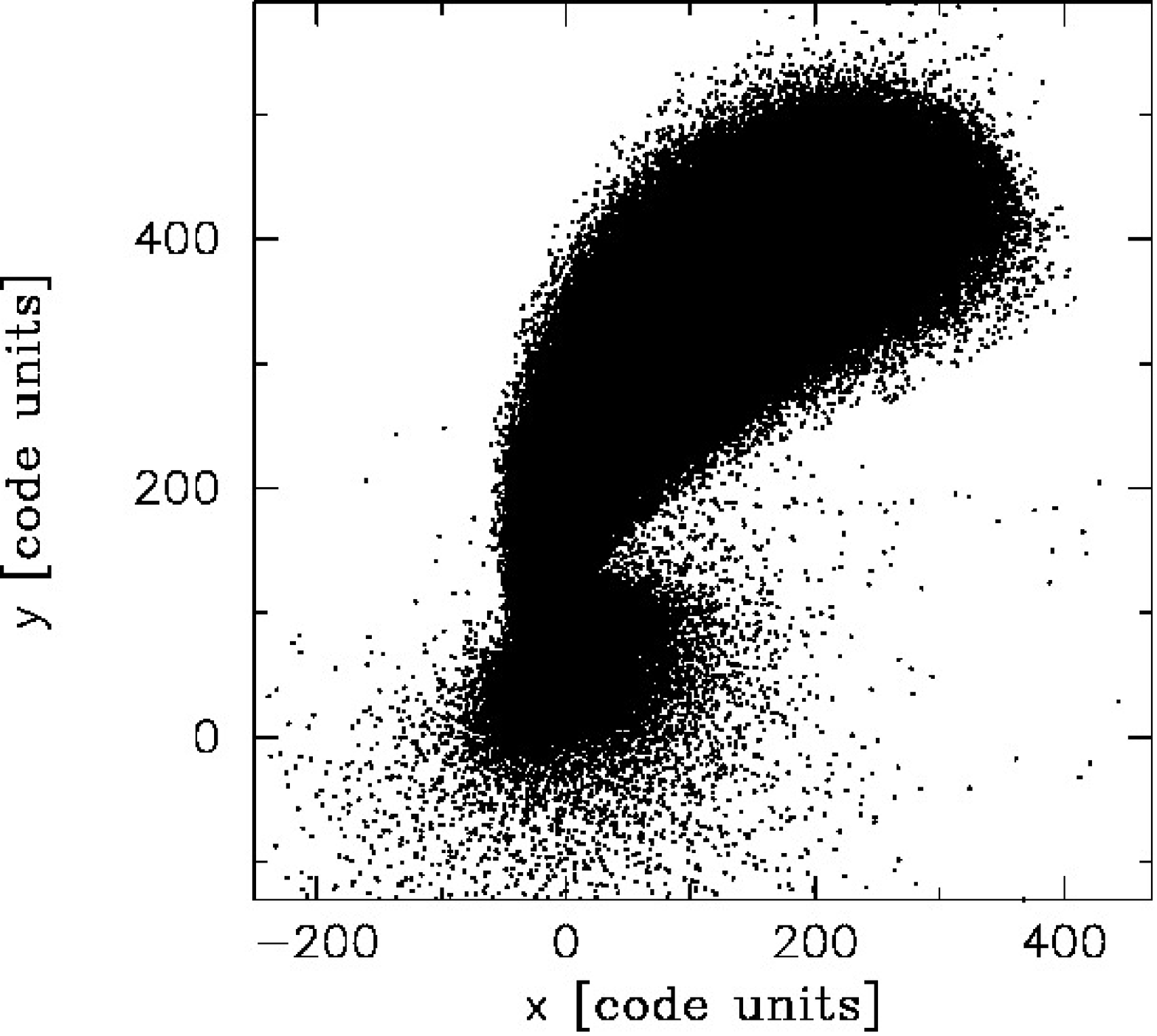}
\caption{Snapshot from a tidal disruption of a white dwarf (500677 particles) by an 
intermediate-mass black hole of 1000 \msun. The large range of smoothing lengths, five orders of magnitude,
make the force calculation challenging.}
\label{fig:imbh_splash}
\end{minipage}
\end{figure*}

Motivated by possible future applications of our RCB tree code, we chose
three sets of initial conditions: particles distributed in a sphere according to a quasi-random
Sobol distribution \citep{press92}, see Fig.~\ref{fig:sobol_splash}, and two snapshots from SPH simulations:
a white dwarf -- white dwarf (WD--WD)  collision with 602506 particles \citep{rosswog09c}, 
see Fig.~\ref{fig:wdwd_splash}, and a WD (500677 particles) that has been tidally disrupted by an 
intermediate-mass black hole (IMBH) of 1000 \msun $\:$ \citep{rosswog09a}, see Fig.~\ref{fig:imbh_splash}.
The Sobol sequence yields particles that homogeneously fill the allowed space (here a sphere), but its 
quasi-randomness makes it a `worst case scenario' for efficient memory access. The other two 
scenarios are close to the future applications of our code and they are challenging in terms of 
a complicated geometry with a very large spectrum of particle densities and smoothing lengths 
and in the sense that the particles are scattered across memory, i.e. spatially nearby particles 
are {\em not} close in terms of memory location.\\
As explained above, our tree will become an ingredient of several codes
which will differ in their time integration methods. Therefore, we only present tests
of speed and force accuracy calculations for static particle distribution snapshots.
We aim for an average relative force error of $\approx 0.1\%$, but also monitor the force 
errors of all particles to make sure that even the most extreme outliers remain at an acceptable 
accuracy. This is crucial since among our intended applications are neutron stars
which, as a result of their nearly incompressibe matter equation of state at high 
density, exhibit very sharp surfaces. Particles with large errors in such a surface 
region can be disastrous for the whole simulation.\\
For all tests we compiled the serial code with the Intel Fortran 9.1 compiler, using the \texttt{-O3} 
optimisation flag and the double precision flag \texttt{-real-size 64}. The code was executed on 
an Intel Xeon E5420 processor running at 2.50 GHz, with 6 MB of L2 cache and 8 GB of RAM.

\subsection{Accuracy}\label{subsec:accuracy}
\begin{figure}
\centering
\includegraphics[clip,width=84mm]{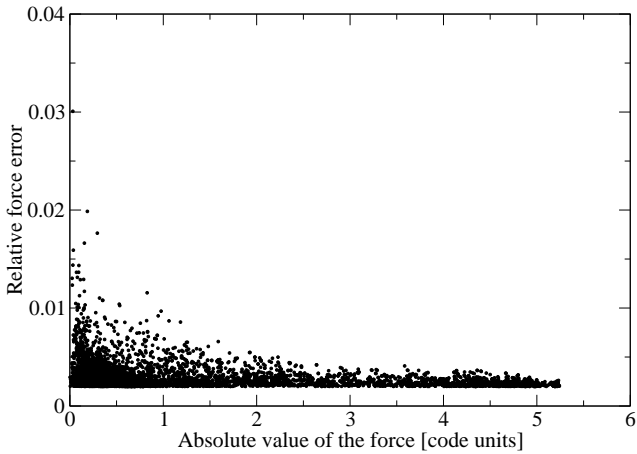}
\caption{The relative error $\varepsilon=|\delta\mathbfit{a}|/|\mathbfit{a}|$ 
plotted against the absolute value of the force, $|\mathbfit{a}|$. 
Accuracies of $\varepsilon < 0.2\%$ are obtained for particles across the full force
spectrum (for over $99\%$ of all particles). These particles are {\em not} shown here,
instead, we focus in this plot on the outliers. Only particles with very small forces 
exhibit force errors in excess of 1\%.  This figure has been obtained for the particle distribution of a
WD--WD encounter ($N=602506$, $\theta=0.7$). All other cases yield very similar
distributions.}
\label{fig:errvsforce}
\end{figure}

As a first test we explored the sensitivity of the force accuracy to the parameter $\theta$.
The cumulative distribution of relative errors in the acceleration, $\varepsilon_i=
|\delta\,\mathbfit{a}_i|/|\mathbfit{a}_i|$, was investigated for $\theta$ in 
a range from 0.4 to 1. Our RCB tree results are compared against direct, 
kernel-smoothed summations according to  
Eq.~(\ref{eq:direct_kernel_sum}) which are correct to double precision accuracy 
(15 significant digits).
\begin{figure}
\centering
\includegraphics[clip,width=84mm]{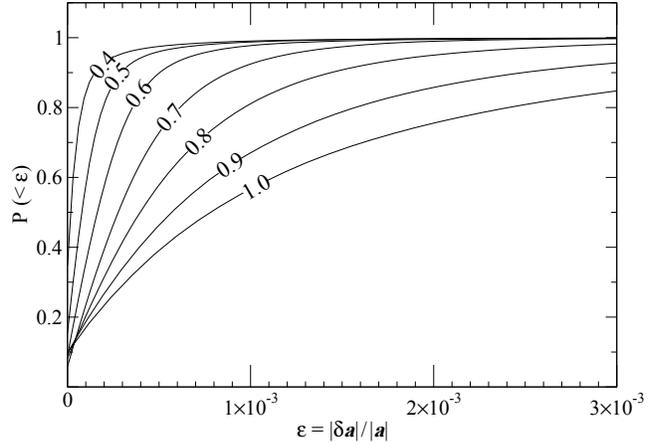}
\caption{Cumulative error curves for $\theta=0.4, 0.5, 0.6, 0.7, 0.8, 0.9$ and $1.0$, 
showing the fraction $P(<\varepsilon)$ of particles with relative errors $|\delta\mathbfit{a}|/|\mathbfit{a}|$ 
smaller than the value $\varepsilon$ given on the $x$ axis. 
This figure has been obtained for the particle distribution of a
WD--WD encounter ($N=602506$, $\theta=0.7$). All other cases yield very similar
distributions.}
\label{fig:accuracyplot}
\end{figure}
The cumulative error distributions are plotted in Fig.~\ref{fig:accuracyplot}. 
For $\theta=0.7$, 99\% of the particles have relative errors less than 0.2\%. 
Fig.~\ref{fig:errvsforce} shows in detail the relation between the
relative force errors $\varepsilon_i$ and the absolute values of the force, $|\mathbfit{a}_i|$,
in a typical simulation. Since the plots for our test cases are quantitatively very similar,
we only present the results for the WD--WD encounter. Also, since the particles with
$\varepsilon < 0.2\%$ are uniformly distributed across the entire force spectrum, we do
not plot them. The plot demonstrates that all the particles with large relative errors
only experience a very weak total gravitational force.\\
The above accuracy is acceptable for most astrophysical applications, therefore we make $\theta=0.7$
our default value.  We find that the accuracy plots for all three particle
distributions are virtually identical, therefore we display only the results 
for the WD--WD encounter. This robustness with respect to the geometry
of the particle distribution underlines the versatility of our approach and is a 
highly desired property for an astrophysical simulation algorithm. In 
all of the cases, the very few particles with relative errors above 0.2\% feel
an essentially zero net force. Formally, of course, this relative error can 
diverge (for exactly  $|\mathbfit{a}_i|=0$), but in practice this does not have 
any influence since the particles just do not move during a time step. \\
\cite{salmon94} showed that the conventional geometric MAC, see Eq.~(\ref{eq:MAC}), 
can introduce unbounded errors in certain pathological cases (see their Appendix A for 
the `detonating galaxies' scenario).  In our RCB tree, these situations are avoided in 
two ways. On the one hand, the CMS described in Sec. \ref{subsec:build} protects against 
having very distant particles in the same cell. On the other hand, since the particle content
of each reachable cell is summed up directly and independent of the MAC, we are 
protected against potentially unbounded errors from particles that get very close to the 
sink but whose cells are, for some reason, still accepted by the MAC.

\begin{table}
\caption{Scaling of the RCB tree with the average number of particles per ll-cell, $\overline{N}_{ll}$. 
The following abbreviations are used: L: number of tree levels; P$_{<1 \%}$: fraction of particles
with an error smaller than 1 \%; TB: tree build, N: neighbour search, NG: near-field gravity, 
FG: far-field gravity, G; total gravity. Times are measured in seconds.}
\label{tbl:npcmax}

\begin{tabular}{@{}rcccrrrr}
  \hline
$\overline{N}_{ll}$  & L & P$_{<1 \%}$ & TB [s] & N [s] & NG 
[s] & FG [s] & G[s] \\
  \hline
    1 & 19 & 0.9998 & 0.76 & 23.54 & 6.14   & 99.26 & 171.47  \\
    3 & 18 & 0.9998 & 0.51 & 10.90 & 7.15   & 39.76 & 75.52   \\
    6 & 17 & 0.9998 & 0.36 & 6.67  & 10.16  & 16.12 & 37.83   \\
   12 & 16 & 0.9998 & 0.29 & 5.86  & 14.66  & 6.66  & 25.90   \\
   24 & 15 & 0.9997 & 0.25 & 7.07  & 22.02  & 2.71  & 26.67   \\
   48 & 14 & 0.9997 & 0.24 & 10.56 & 36.22  & 1.02  & 38.13   \\
   64 & 13 & 0.9996 & 0.22 & 14.97 & 56.90  & 0.50  & 57.88   \\
  128 & 12 & 0.9997 & 0.19 & 26.12 & 101.51 & 0.18  & 102.04  \\
  \hline
\end{tabular}

\medskip
The test was performed on a spherical Sobol distribution with 500 000 particles and smoothing lengths 
chosen so that the average number of neighbours was 105. With our standard value for the accuracy 
parameter, $\theta=0.7$, the average relative error $\bar{\varepsilon}\approx 0.1\%$ for all test cases.
Particles with errors larger than $1\%$ are essentially force free (their force is about five orders of magnitude
smaller than the average force). In an extensive set of tests we always found optimal results for
$\overline{N}_{ll}\approx 12$, regardless of the number of particles or their distribution.
\end{table}

\subsection{Performance} 

\subsubsection{Optimizing the tree depth}
We experimentally optimize the depth of our RCB tree, which is
determined by the average number of particles per ll-cell, $\overline{N}_{ll}$. Sample results 
for a spherical Sobol distribution with $N=500000$ are presented in Table \ref{tbl:npcmax}.
As expected, the tree build becomes faster as the height of the tree
decreases, but in none of the cases the tree build takes more than $1\%$ of the total
computing time. The neighbour search and the gravity calculations depend strongly on the height of
the tree: larger ll-cells mean a shallower tree and thus faster tree walks and far-gravity, 
while the near-gravity neighbour candidate lists become longer and increasingly more 
expensive to evaluate. The balance between the two regimes was empirically found 
at $\overline{N}_{ll}\approx12$ regardless of the number of particles or the complexity of their 
spatial distribution. The performance is rather robust against substantial changes in this number, 
e.g. doubling this value to 24 yields essentially the same performance.
All of our subsequent test calculations were obtained with $\overline{N}_{ll} = 12$.

\subsubsection{Comparison with the Press tree} 
We chose to compare the RCB tree with a commonly used binary tree that is due to 
Press \citep{press86,benz89}, mainly because we have used this tree
for years and because this is the tree that our new RCB tree will replace. The version 
used in this comparison has been tuned by collecting quantities that are frequently used together 
into common arrays \citep{rosswog07c}. This has improved the overall performance for completely 
unsorted particle arrays by nearly a factor of three with respect to the original version. We
have chosen the accuracy parameters of the two trees so that we obtain overall very similar 
accuracies, an average relative error of $\bar{\varepsilon}\approx 0.1\%$ and a maximum error 
$\varepsilon_{\mathrm{max}}\la 1\%$, but with the RCB tree being slightly more restrictive and 
producing slightly higher accuracies. This motivated our choice of $\theta_{\rm Press}= 0.5$ for the
Press and $\theta_{\rm RCB}= 0.7$ for our RCB tree. The larger $\theta$-value for the RCB tree has two 
explanations: the Press tree has tight bounds by construction \citep{anderson93} and thus 
exhibits physically smaller nodes, and the MAC in our RCB tree is more conservative in the sense
that it chooses the longest cell edge as a measure of the cell size, see Eq.~(\ref{eq:MAC}).\\

\noindent{\it Tree build}\\ 
In a first test we compare the performance during the tree building phase. Due to the more
complicated algorithm for identifying mutual nearest neighbours, building the Press tree 
from scratch is quite time consuming. In fact, from all the trees frequenty used in astrophysics,
the Press tree is probably the most expensive one to build.
\begin{figure}
\centering
\includegraphics[clip,width=84mm]{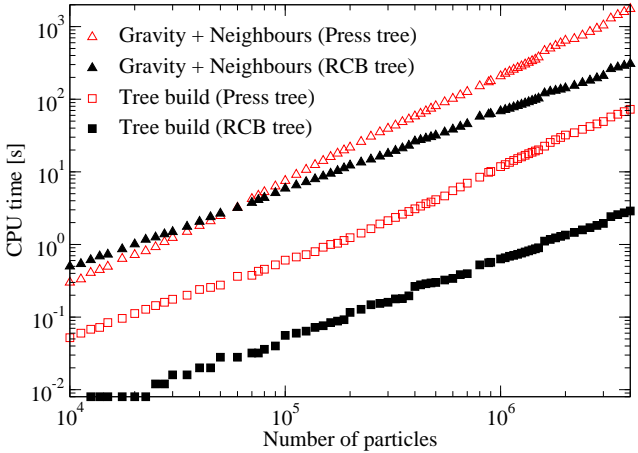}
\caption{Comparison between the Press and the RCB tree in terms of the tree building time 
and of the cumulative time required for neighbour search and gravity calculations. The particle 
distributions for this test were spherical Sobol distributions with up to 4 million particles. 
The smoothing lengths were chosen such that the average neighbour number was 105. The plot 
shows a cross-over point for the gravity + neighbours timings at $N \approx 50000$: for very 
low particle numbers, the complex infrastructure of the RCB tree renders it slower (up to a 
factor of two) than the Press tree.}
\label{fig:TimingsRCBPress}
\end{figure}
As test cases we chose spherical Sobol distributions
with smoothing lengths so that the average neighbour number was 105 (this is conservative for
the gravity calculation with our RCB tree algorithm since it uses large particle numbers 
in the direct, near-gravity summation and the results would be even stronger in favour
of the RCB tree for smaller neighbour numbers). The results of this test are presented in 
Fig.~\ref{fig:TimingsRCBPress} with filled (open) squares for the RCB (Press) tree. Here, 
the RCB tree turns out to be substantially faster than the Press tree, for $4 \times 10^6$ 
particles (the maximum we could afford for our version of the Press tree) on one processor 
our RCB tree build is already $\sim 25$ times faster than the Press tree with 
increasing discrepancy for larger particle numbers.\\

\noindent{\it Neighbour search and gravity}\\
The second performance measure is the sum of the times for  neighbour search and 
gravity. The reason for summing them up is that our version of the Press tree performs 
both operations in the same loop for increased performance. We decided to keep the RCB 
tree code flexible and separate the two operations in different subroutines that can be 
called independently. One would therefore expect the RCB code to perform even better
if one was willing to sacrifice this flexibility and to calculate both in a single loop.
The RCB tree outperforms the Press tree in this test already at $N\approx 50 000$, see
the filled triangles for the RCB and the open triangles for the Press tree. 
Up to this point its relatively complex infrastructure makes the RCB tree slower (up to a 
factor of two, for these particle numbers corresponding to fractions of a second on one 
processor). Near $4 \times 10^6$ particles,  however, RCB neighbour search and 
gravity are faster by about a factor of 6 with larger discrepancies for increasing particle 
numbers. In this and the subsequent plot one notices little `glitches' in the RCB 
curves. They occur whenever the height of the tree changes, since tree walks and 
far-field calculations, being performed `per ll-cell' rather than `per particle', only 
depend on the number of ll-cells and not directly on the number of particles.

\subsubsection{Behaviour for large $N$}
In order to test the robustness and the scaling 
behaviour of our RCB tree we computed, on one processor, the self-gravity of Sobol 
distributions with increasingly larger particle numbers, ending only at $10^{8}$ particles. 
Successfully running such a simulation on a machine with only 8 GB of RAM gives an upper limit
on the memory consumption of our code of approximately 86 bytes per particle. This includes
not only the arrays that store particle properties, but also the ones that store the nodes, 
the temporary variables, the subroutines, the files read in memory, and 
every other bit of RAM that our code uses.\\
\begin{figure}
\includegraphics[clip,width=84mm]{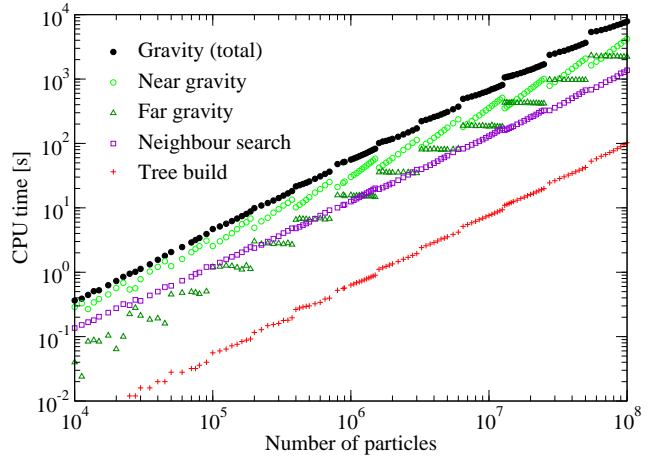}
\caption{Comparative timings of the RCB tree code components (tree build, neighbour search 
and gravity calculations) for spherical Sobol distributions with up to $10^{8}$ particles 
and $\sim 105$ neighbours per particle. Since the far-field gravity calculation does not depend
on the particle number, but only on the tree depth the `far gravity' in the plot remains 
essentially constant until the number of tree levels is incremented.
The tree build always takes less than 1\% of the execution time, the gravity calculation typically 
takes 80-90\% percent, and the neighbour search occupies the remaining 10-20\%.}
\label{fig:TimingsRCBLogarithmic}
\end{figure}
In Fig.~\ref{fig:TimingsRCBLogarithmic} we compare the time spent for different operations. 
The tree build becomes completely negligible as the number of particles increases 
(even in the worst case it never surpasses 1\% of the total execution time). The tree build in the Press tree
has been one of the most expensive components, see also \cite{nelson09}, and has been 
avoided whenever possible, usually by `revising' the tree and amortising the cost of 
the tree build over a number of time steps. This, however, can become difficult: 
simulations involving black holes where particles are frequently absorbed at the horizon 
require a frequent tree build (if they orbit safely inside the horizon they can be temporarily
stored in a special list and be removed only later, see \cite{rosswog05a}, however at a 
substantial bookkeeping effort). Furthermore, since the tree build is usually the 
component of the code with the worst parallel scaling, having a fast tree build algorithm 
becomes crucial when parallelising the code. Therefore, the tree build is, together with the 
scaling behaviour close to $\mathcal{O}(N)$, one of the strongest points of the RCB tree.
The time for neighbour search takes 10-20\%, the remaining 80-90\% are invested for the 
gravitational forces.\\
\begin{figure}
\centering
\includegraphics[clip,width=84mm]{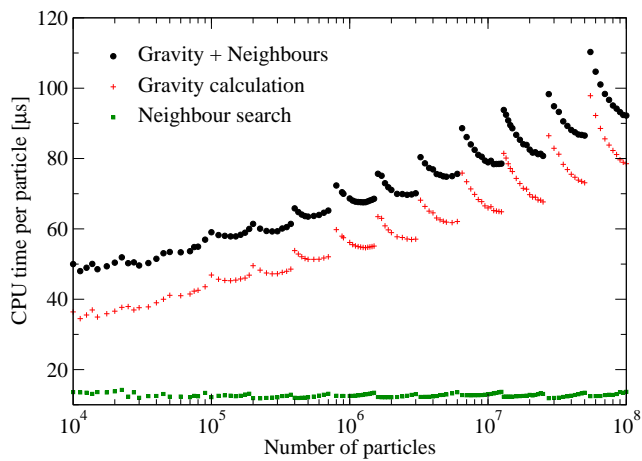}
\caption{CPU time spent per particle in neighbour search, gravity calculations, and 
cumulative neighbour search + gravity calculations for spherical Sobol distributions 
with up to $10^{8}$ particles and $\sim 105$ neighbours per particle. For a given number
of levels the RCB tree scales slightly better than $\mathcal{O}(N)$, and
the execution time only increases by a constant (the plot is shifted upwards) 
when another level is added to the tree. Overall, the execution time per particle increases
much slower than the $\log(N)$ expected for standard tree codes. When increasing the
particle number by a factor of $10^4$ the execution time per particle increases merely
by a factor of two. }
\label{fig:TimingsRCBPerPart}
\end{figure}
In Fig. \ref{fig:TimingsRCBPerPart} we display the 
CPU time spent per particle in different parts of the code. This figure shows the  
$\mathcal{O}(N)$ scaling of the neighbour search, even at $10^{8}$ particles (the time 
spent searching the neighbours of one particle is always $\approx 13$ $\umu\mathrm{s}$). 
At a given tree depth the gravity calculation scales slightly better than $\mathcal{O}(N)$ 
for a given height of the tree (i.e., as more particles are added to the same cells, the 
CPU time per particle drops). This is partly an effect of the grouped tree traversal: since one tree 
walk and one `far force' evaluation are performed per ll-cell, these two components of 
the code do not depend on the number of particles, but only on the height of the tree.
This makes filling an ll-cell with more and more particles increasingly more efficient in terms of
tree-traversals and far-gravity calculations (such behaviour can occur in any tree code, 
since adding one more level to a tree always increases the time needed by the tree walk, 
regardless of whether it is done per particle or per ll-cell). The near-gravity calculation, 
however, becomes more expensive since a larger number of particles is summed up directly.  
The total time per particle spent with tree operations is increasing somewhat, but at a 
much slower pace than the $\log(N)$-behavior encountered for standard tree codes. When 
increasing the total particle number by a factor of $10^4$ the time per particle merely 
increases by a factor of two.

\subsubsection{OpenMP parallelisation}
\begin{figure}
\centering
\includegraphics[clip,width=84mm]{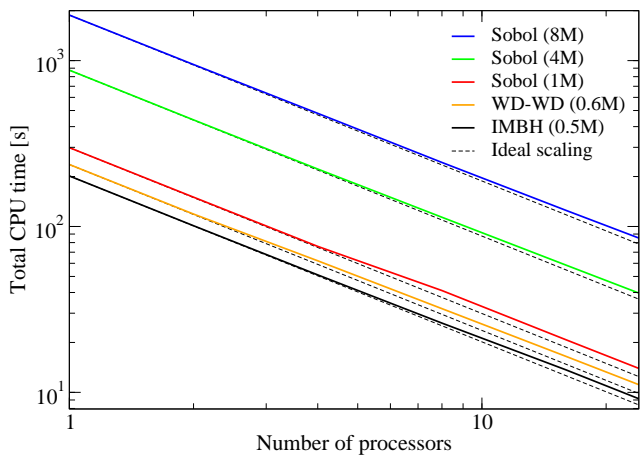}
\caption{Scaling of the total CPU time (tree build + neighbour search + gravity calculations) 
with the number of processors. This plot also shows, with dashed lines, the ideal
 scaling for each of the five test cases. Not only are the results very close to the 
ideal scaling, but they improve as the
number of particles $N$ increases.}
\label{fig:scalingOMP}
\end{figure}

The parallel code was compiled and executed on an SGI Altix 3700Bx2 machine with 24 Itanium2 
processors (Madison9M) running at 1.6 GHz, with 6 MB of L2 cache and 96 GB of shared RAM. We 
tested the scaling of the RCB tree code  with both the number of processors and particles. 
We chose three Sobol distributions (with 1, 4 and 8 million particles, respectively) 
and the two snapshot tests described above; the corresponding results are presented in 
Fig.~\ref{fig:scalingOMP}. The results obtained in this test are rather robust against changes in  the
particle distribution and even particle number. In all cases a speedup of $>21$ was obtained on
24 processors.

\section{Conclusions} 
We have introduced a new tree for neighbour search and gravity  that is based on recursive 
coordinate bisection. This binary RCB tree is only built down to an optimal depth so that there
are still about 12 particles left per lowest-level cell. This property makes the tree fast (since 
some operations are only performed per lowest-level cell rather than per particle)
and  memory efficient since fewer aggregated quantities need to be stored for such a shallow
tree.
Gravity is split into a near- and far-field component; the first is calculated via a direct, kernel-smoothed
summation while the latter uses a `cell-cell interaction' based on Cartesian multipole and Taylor expansion.\\
We have compared the performance of our RCB tree against that of the `Press tree' that we had used
earlier on various occasions. As expected for such a `top-down' tree, the tree building phase is subtantially
faster for RCB tree. At four million particles the tree build is faster by about a factor of 25, neighbour
search and gravity (at slightly higher accuracy for the RCB tree) are faster by a factor of six. These ratios
become even more favorable for the RCB tree with increasing particle numbers. The code was 
tested for up to $10^{8}$ particles on a single processor, showing very good scaling behaviour close to 
$\mathcal{O}(N)$  and a low memory consumption.

\section*{Acknowledgements}
\balance
We would like to acknowledge useful discussions with Ivo Kabadshow from Forschungszentrum J\"ulich.
This work was supported by  Deutsche Forschungsgemeinschaft under grant number RO-3399/5-1.
It has profited from visits to Oxford which were supported by a DAAD grant  `Projektbezogener 
Personenaustausch mit Gro{\ss}britannien' under grant number 313-ARC-XXIII-Ik and from visits
to La Scuola Internazionale Superiore di Studi Avanzati (SISSA), Trieste, Italy. It is a pleasure to acknowledge
in particular the hospitality of John Miller (Oxford, Trieste).
We gratefully acknowledge the hospitality of the University of Basel (E.G.)
and the University of Queensland in Brisbane (S.R.) during the time when 
this paper was finalized. S.R.'s stay in Brisbane was supported by the 
DFG by a grant to initiate and intensify bilateral collaboration.

\bibliographystyle{mn2e}
\bibliography{rcbpaper.bib}

\bsp
\label{lastpage}

\end{document}